\documentclass{emulateapj} \usepackage{apjfonts}

\newcommand{\be}{\begin{equation}}
\newcommand{\ee}{\end{equation}}

\def\ltsima{$\; \buildrel < \over \sim \;$}
\def\simlt{\lower.5ex\hbox{\ltsima}}
\def\gtsima{$\; \buildrel > \over \sim \;$}
\def\simgt{\lower.5ex\hbox{\gtsima}}

\begin{document}

\title{Formation of A Spiral Galaxy in A Major Merger}

\author{Volker Springel\altaffilmark{1} \& Lars
Hernquist\altaffilmark{2}}

\altaffiltext{1} {Max-Planck-Institut f\"{u}r Astrophysik,
Karl-Schwarzschild-Strasse 1, 85740 Garching, Germany;
volker@mpa-garching.mpg.de}

\altaffiltext{2} {Harvard-Smithsonian Center for Astrophysics, 60
Garden St., Cambridge, MA 02138; lars@cfa.harvard.edu}
 
\begin{abstract}
We use numerical simulations to examine the structure of merger
remnants resulting from collisions of gas-rich spiral galaxies.  When
the gas fraction of the progenitors is small, the remnants
structurally and kinematically resemble elliptical galaxies, in
agreement with earlier work.  However, if the progenitor disks are
gas-dominated, new types of outcomes are possible. In fact, we show
that a prominent disk may survive in certain cases.  To illustrate
this scenario, we analyze an extreme example with progenitor galaxies
consisting of dark matter halos, pure gas disks, and no bulges, as
might be appropriate for mergers at high redshifts.  While rapid star
formation triggered by tidal torques during the merger forms a
central, rotating bulge in the remnant, not all the gas is consumed in
the burst. The remaining gas cools very quickly and settles into an
extended star-forming disk, yielding an object similar to a spiral
galaxy, and {\em not} an early type galaxy.  This is contrary to the
usual view that major mergers invariably destroy disks.  The
morphological evolution of galaxies can therefore be more complicated
than often assumed, and in particular, theoretical constraints based
on the fragility of spiral disks need to be reevaluated.

\end{abstract}


\section{Introduction}
\label{intro}

Mergers and interactions between galaxies are an essential ingredient to
galaxy formation and evolution.  The gravitational tidal forces associated
with this process can explain the morphological characteristics of peculiar
galaxies \citep{Toomre1972}, and it is believed that mergers can trigger the
elevated levels of star formation seen in ultraluminous infrared galaxies
\citep{Sanders1988,Melnick1990}.  Various observations suggest that quasars,
radio galaxies, and active galactic nuclei (AGN) are formed in mergers
\citep[for reviews, see e.g.][]{BarnesHernquist1992,Jogee2004}.
Moreover, according to
hierarchical models of structure formation \citep{White1978}, it is expected
that galaxies grow with time through mergers.

\citet{Toomre1977} was among the first to recognize that mergers can drive the
evolution of galaxy types by transforming disks into objects that resemble
ellipticals.  This idea was examined numerically by
\citet{Barnes1988,Barnes1992} and \citet{Hernquist1992,Hernquist1993a} in the
limit where dissipational effects arising from gas dynamics are negligible,
and it was shown that mergers involving equal-mass galaxies (i.e.~``major''
mergers) do indeed yield remnants with properties similar to those of observed
ellipticals.  Simulations including gas dynamics and simple prescriptions for
star formation and feedback have further demonstrated that major mergers can
drive gas to the center of a remnant \citep{BarnesHernquist1991,BarnesHernquist1996}, triggering starbursts
with intensities similar to those of observed ultraluminous infrared galaxies
\citep{Mihos1996}.

While major mergers are the most striking examples of galaxy
collisions, ``minor'' mergers between galaxies of different masses are
probably at least an order of magnitude more frequent
\citep{Ostriker1975, Toomre1981}.  Simulations have shown that
dissipationless minor mergers between spiral galaxies and smaller
companions can cause significant perturbations to disks through
dynamical heating \citep{Quinn1986, Quinn1993,
Walker1996,Velazquez1999}.  Even with large mass ratios $\sim 10 : 1$,
the damage can be severe because disks of spirals are dynamically
cold.  This conclusion is unaffected by dissipation when the disks
contain a small fraction ($\sim 10\%$) of their mass in gas
\citep{Hernquist1989a, Hernquist1995}.

Together, the results for both major and minor mergers indicate that
galaxy collisions are problematic for the long-term survivability of
disks.  \citet{Toth1992} used this notion to constrain the
cosmological merger rate by arguing that disks like that of the Milky
Way would not survive to the present day if they were constantly
bombarded by smaller companions, posing a challenge for hierarchical
galaxy formation.

However, the argument put forward by \citet{Toth1992} was based on theoretical
models in which the interstellar gas was at most a small fraction of the disk
mass.  Previous numerical studies of minor mergers were restricted to cases
where the gas fraction was low because of gravitational instabilities in
isolated galaxies with cold, gas-dominated disks.  The simulations of
\citet{Hernquist1989a} and \citet{Hernquist1995}, for example, employed disks
with $10\%$ gas because these authors described the interstellar medium (ISM)
using an isothermal equation of state and they included only a minimal form of
kinetic feedback from massive stars \citep{Mihos1994a}.  This approach
produces unstable models for much larger gas fractions.

Recently, \citet{Springel2004c} have developed a method for constructing
stable disks with arbitrary ratios of gas to stellar mass.  Their approach
employs a sub-resolution, multiphase model of the ISM that captures the impact
of star formation and supernova feedback on resolved scales
\citep{Springel2003a}.  While the ``microscopic'' structure of the ISM is not
followed in detail in this methodology, the coarse-graining that is used to
form a ``macroscopic'' representation of star formation describes the
consequences of feedback in a simple and physical manner through an effective
equation of state (EOS) for the star-forming gas.  In this picture, feedback
from star formation pressurizes the gas so that the effective EOS is stiffer
at high densities than if the gas had been isothermal, stabilizing disks
against fragmentation.

\begin{figure}
\begin{center}
\vspace*{-0.2cm}\resizebox{8.5cm}{!}{\includegraphics{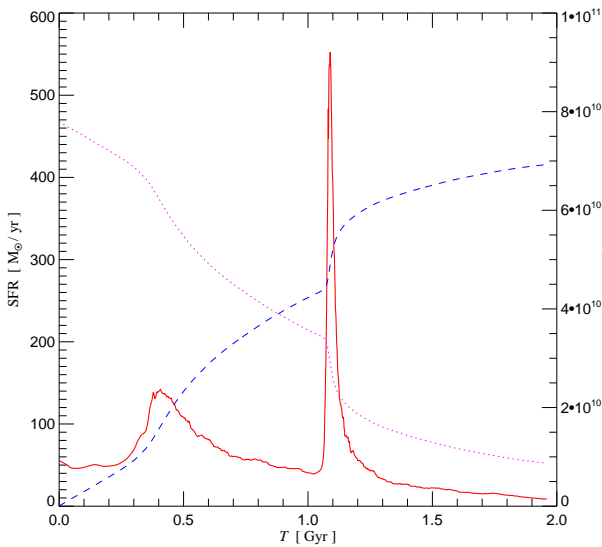}}\vspace*{-0.1cm}%
\end{center}
\caption{Evolution of the star formation rate and gas and stellar mass
  in a major merger of two disk galaxies without bulges.  The disks of
  the galaxies initially consisted entirely of gas.  The solid line
  shows the evolution of the star formation rate (left axis), 
  while the dashed and dotted lines give the evolution of the total stellar and
  gas mass (right axis) of the galaxy pair.}
\label{fig:sfr88}
\end{figure}

In what follows, we use the \citet{Springel2004c} procedure to simulate major
mergers between disk galaxies with large gas fractions.  We demonstrate that a
new type of outcome is possible when the modeling is extended to cases where
the galaxies are very gas-rich, as might be appropriate for systems at high
redshifts.  In particular, we find that if sufficient gas remains following a
major merger, cooling can quickly reform a disk, yielding a remnant that,
structurally and kinematically, more closely resembles a spiral galaxy than an
elliptical.  This outcome is contrary to the usual view that mergers
invariably destroy disks.  Clearly, the issue of disk survival in a
hierarchically evolving universe needs to be reexamined in the context of our
models.

\begin{figure}[t]
\begin{center}
\vspace*{-0.5cm}\resizebox{4.5cm}{!}{\includegraphics{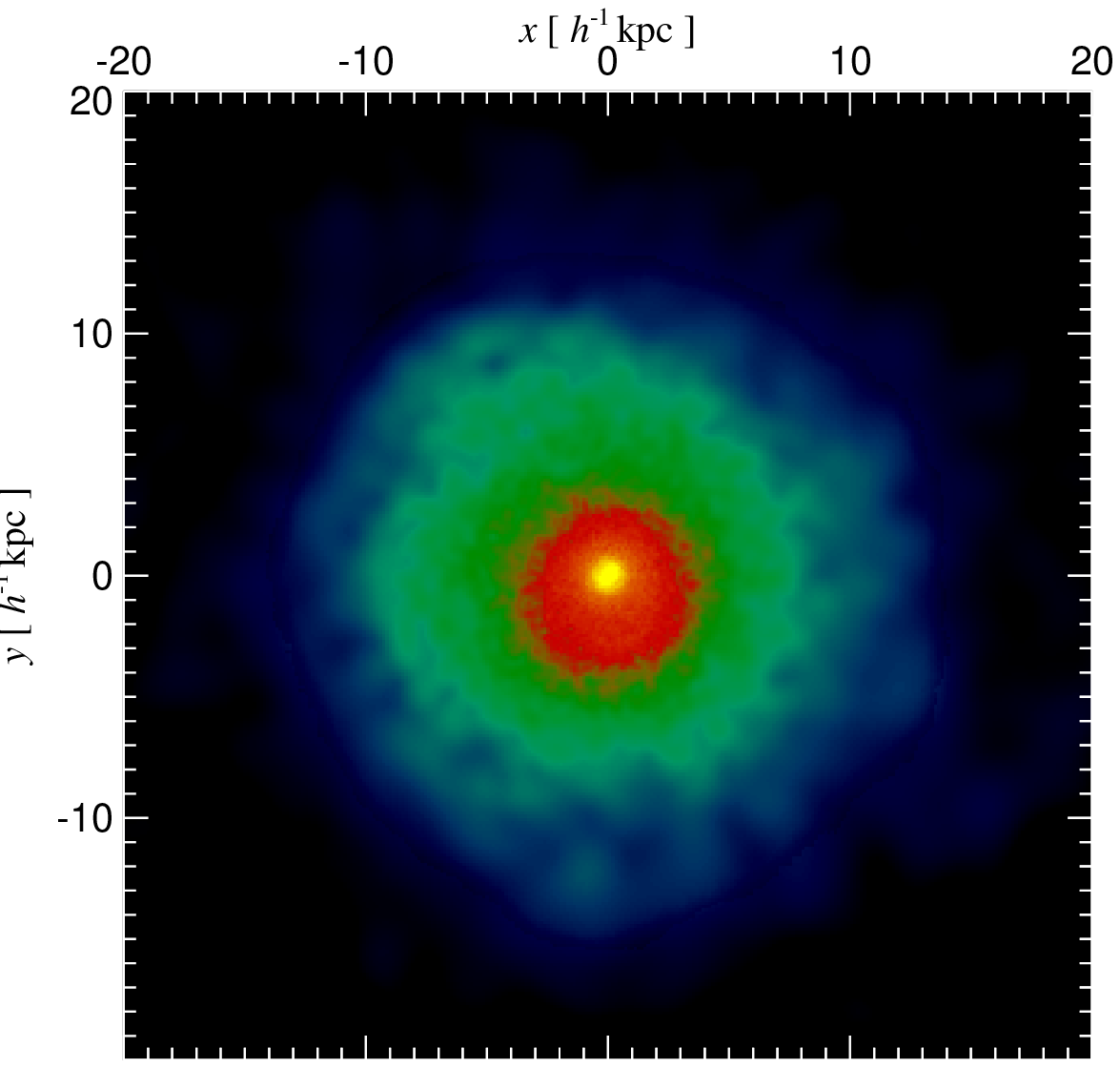}}\hspace*{-0.2cm}%
\resizebox{4.5cm}{!}{\includegraphics{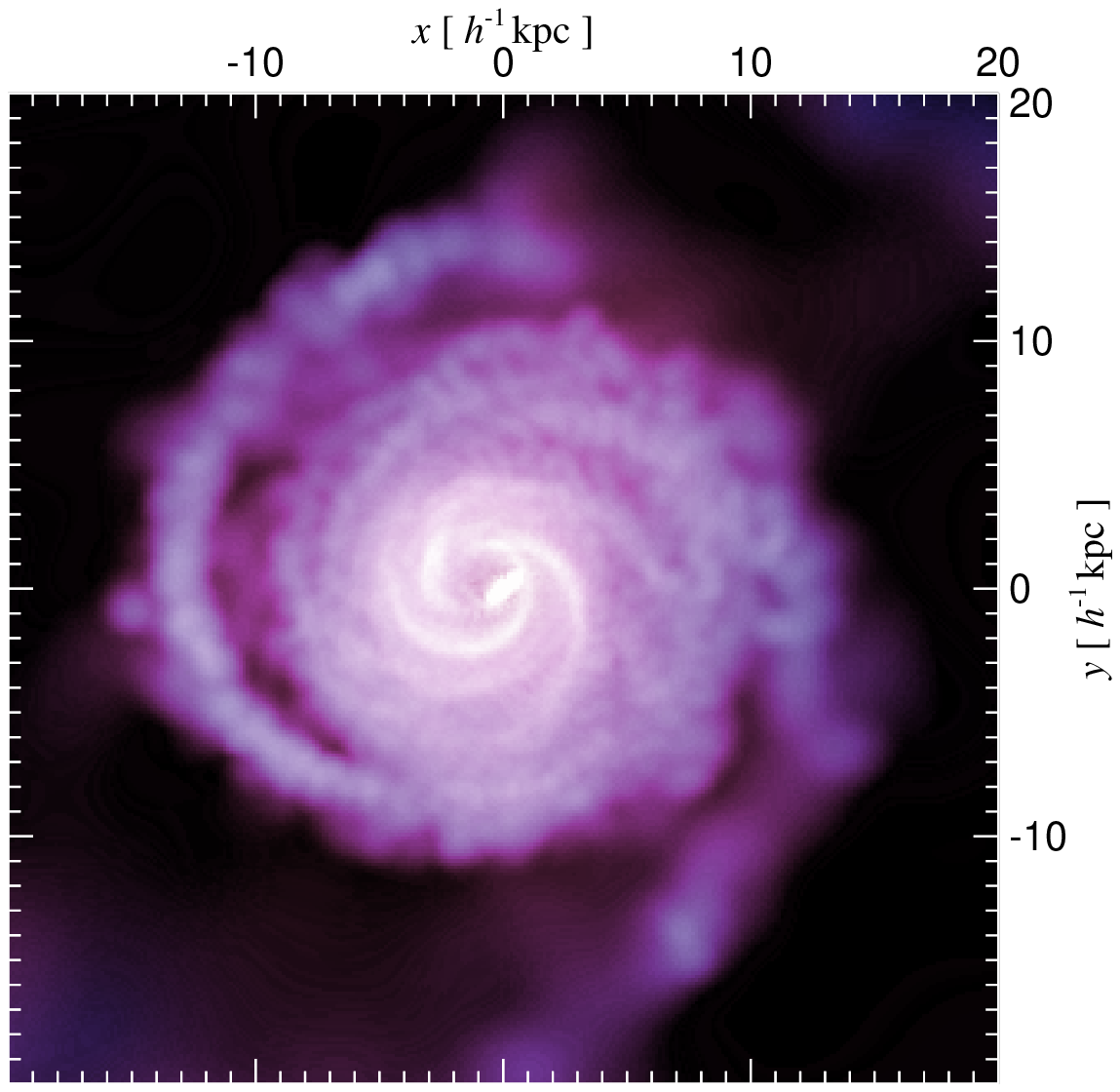}}\vspace*{-0.24cm}\\%
\resizebox{4.5cm}{!}{\includegraphics{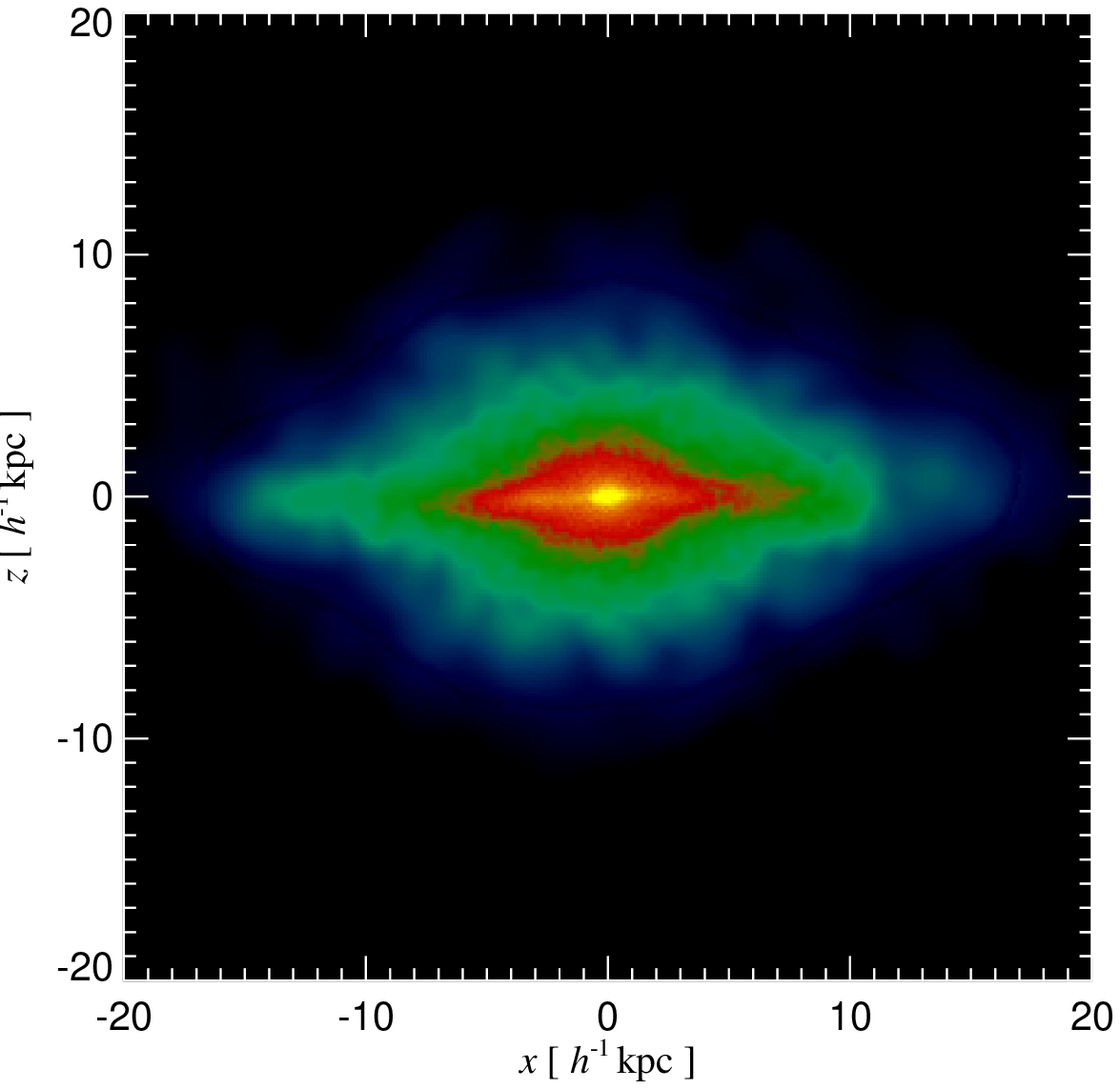}}\hspace*{-0.2cm}%
\resizebox{4.5cm}{!}{\includegraphics{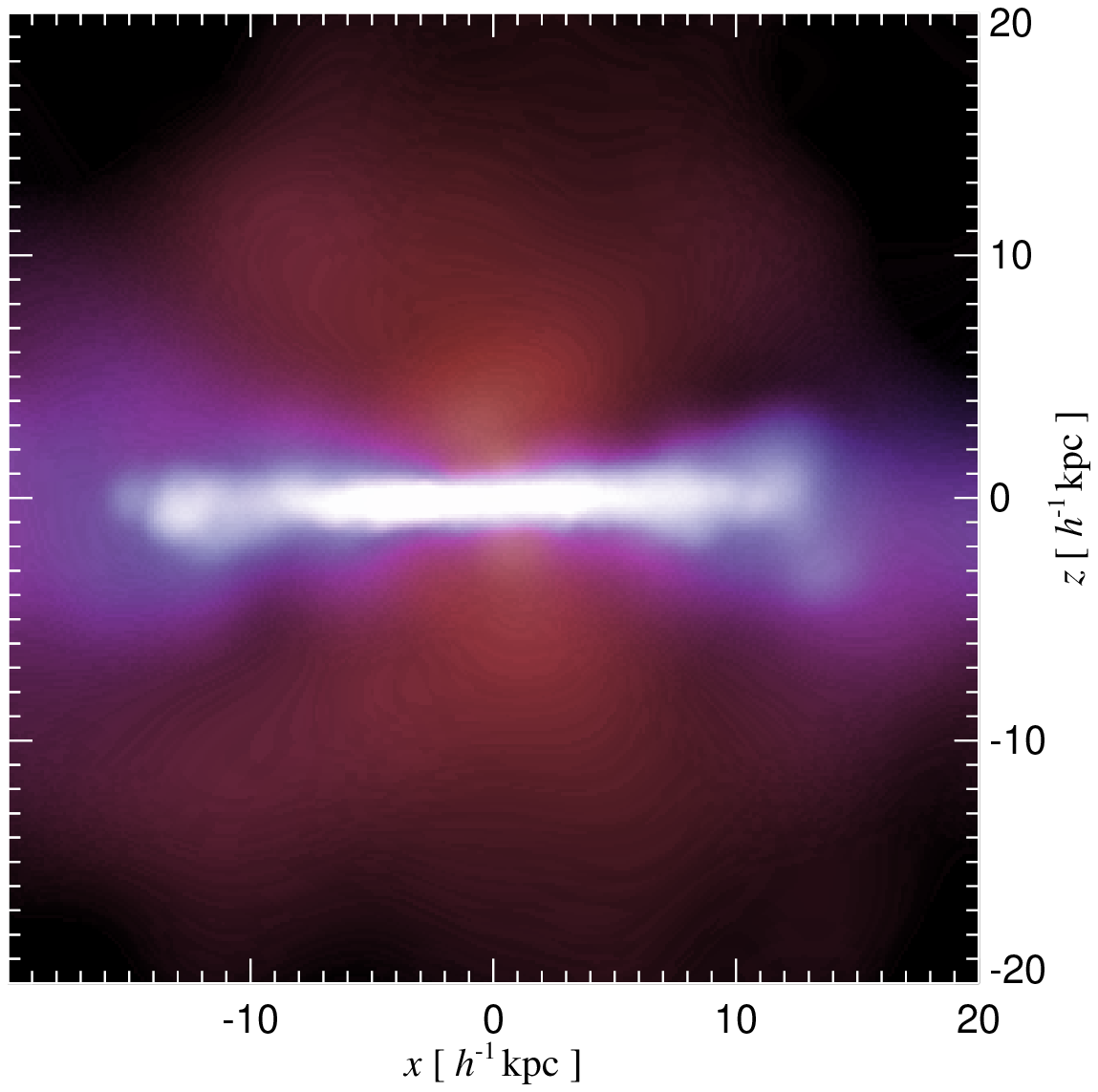}}\vspace*{-0.2cm}\\%
\end{center}
\caption{Distribution of stars (left panels) and gas (right panels)
  following the completion of the merger at a time $t = 1.96$ Gyr, when
  the inner parts of the remnant have relaxed.  The top panels show a
  face-on view of the remnant disk, while the bottom panels are
  edge-on.}
\label{fig:prj88}
\end{figure}

\section{Methodology}
\label{meth}

In \citet{Springel2004c}, we constructed near-equilibrium galaxy
models consisting of dark matter halos, disks of gas and stars, and
optional bulges, using a procedure developed by \citet{Hernquist1993b}
and \citet{Springel2000}, but with a number of refinements.  The dark
matter follows a \citet{Hernquist1990} profile, scaled to match the
inner density distribution of halos found in cosmological simulations
\citep{Navarro1996}.  The disks have exponential surface densities in
both the stars and gas, with the vertical gas profile determined
self-consistently for a particular EOS.  Star formation is described
using a sub-resolution model of the ISM \citep{Springel2003a} to
describe the star-forming gas as a multiphase medium whose structure
is regulated by gas cooling, supernova feedback, and thermal
evaporation of cold clouds.  We have also implemented schemes to
include supernova-driven winds \citep{Springel2003a} and feedback from
black-hole accretion \citep{Springel2004c}, but we ignore these
effects here.

In a parameter study carried out in \citet{Springel2004c}, we have run
a suite of models of both isolated and merging galaxies, varying the
structural properties of the galaxies, the strength of supernova
feedback, the gas fraction of the disks, and the orbit (for
collisions).  In this paper, we focus on one specific case to
illustrate an interesting new type of outcome that is possible when
highly gas-rich disks merge.  The simulation we analyze follows a
major merger of two equal mass galaxies from a prograde, parabolic
orbit.  Initially, each galaxy consisted of a dark matter halo of mass
$M = 9.13\times 10^{11}\,h^{-1}{\rm M}_\odot$, and an exponential disk
of pure gas with mass $M=3.90\times 10^{10}\,h^{-1}{\rm M}_\odot$.
Neither galaxy began with a stellar bulge component.  We choose to
concentrate on this example to simplify the discussion, but our
conclusions are not restricted to the parameters specifying the
galaxies or the orbit, provided that the disks are significantly more
gas-rich than in the earlier work of \citet{BarnesHernquist1991,
BarnesHernquist1996} and \citet{Mihos1996}. For modeling star
formation and feedback, we have used the formalism of
\citet{Springel2003b}, but we softened the equation of state (EOS)
with a factor $q_{\rm EOS}=0.5$ \citep[as discussed
by][]{Springel2004c}, so that the effective pressure at densities
above the star formation threshold is midway between an isothermal EOS
and our full, ``stiff'' multiphase model.  In isolation, the model
galaxies for this choice of $q_{\rm EOS}$ are stable and form stars at
a steady rate owing to the pressurization of the gas from star
formation, even though the disks are pure gas and the galaxies do not
include bulges.

\begin{figure*}[t]
\begin{center}
\vspace*{-0.3cm}\resizebox{12.8cm}{!}{\includegraphics{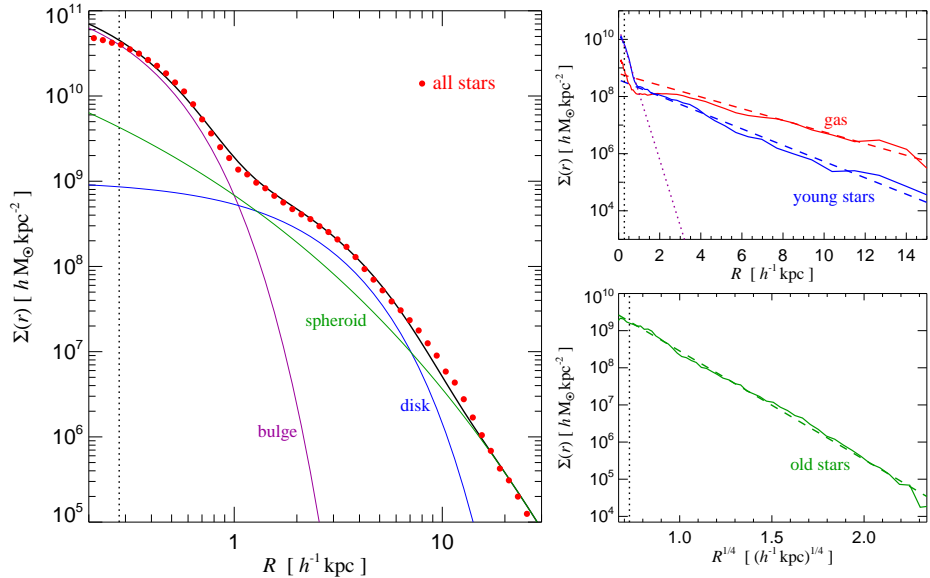}}\vspace*{-0.5cm}
\end{center}%
\caption{Surface mass density distribution of stars following the completion
  of the merger at a time $t = 1.96\,{\rm Gyr}$. The left panel shows the
  measurements for the total stellar density (symbols) together with a fit
  based on three components, a central bulge, an extended spheroid and an
  exponential stellar disk. The scale-lengths of these components are obtained
  by fitting stars of selected ages. `Young' stars are well fitted by an
  exponential (top right panel), while `old' stars in the remnant form a
  $r^{1/4}$ profile (bottom right panel). Dashed vertical lines mark the
  spatial resolution limit of the simulation.}
\label{fig:rem88}
\end{figure*}

We used 120000 particles to represent the dark matter, and 80000 to
represent the gas with SPH particles. We evolved the system over time
with an improved and updated version of the simulation code {\small
GADGET} \citep{Springel2001}, using a fully conservative formulation
of SPH \citep{Springel2002} that maintains strict entropy and energy
conservation even when smoothing lengths vary adaptively.

\section{Results}
\label{results}

In Figure~\ref{fig:sfr88}, we show the evolution of the star formation rate
and the total gas and stellar mass during the merger.  The star formation rate
in each disk prior to the encounter is relatively high, $\sim 20\, {\rm
M}_\odot/{\rm yr}$, because of the high gas content of each galaxy.  When the
galaxies first pass by one another, near time $t\approx 0.4$ Gyr, a moderate
burst of star formation is induced in each disk, owing to the tidal
deformation experienced by each galaxy.  A much stronger burst occurs during
the final coalescence of the two galaxies, near $t\approx 1.1$ Gyr, reaching a
peak amplitude $\simeq 550\, {\rm M}_\odot/{\rm yr}$.  Following the
completion of the merger, the remnant continues to form stars at a declining,
but relatively steady rate $\sim 10\, {\rm M}_\odot/{\rm yr}$.  Star formation
rates at the level of $\approx 500\,{\rm M}_\odot/{\rm yr}$ are similar to
those inferred for systems at high redshift such as Lyman-break galaxies and
SCUBA sources, suggesting that some of these objects may result from mergers
of gas-rich disks.

The evolution shown in Figure~\ref{fig:sfr88} is reminiscent of that
seen by \citet{Mihos1996}, but with several differences, as can be
seen by comparing the results here with the ``Halo/Disk Merger'' in
their Figure 5a.  The starbursts in our new simulations are much more
intense than those found by \citet{Mihos1996} owing to the the larger
gas fractions of our new model galaxies.  In addition, the first
starburst in Figure~\ref{fig:sfr88} at $t\approx 0.4$ Gyr is weaker
than that which follows during the late stages of the merger at
$t\approx 1.1$ Gyr, unlike the behavior found by \citet{Mihos1996} for
their halo/disk mergers.  This difference is a consequence of our
treatment of feedback, which prevents the gas from being strongly
compressed during the first encounter between the galaxies.  This
result demonstrates that the history of star formation predicted for
galaxy mergers is sensitive to assumptions made in describing star
formation.  In the future, it may be possible to constrain these
prescriptions by comparing the simulations with detailed observations,
an approach being pioneered by e.g. \citet{Barnes2004} for the Mice.

As shown in Figure~\ref{fig:sfr88}, about half the gas initially in the
galaxies is converted into stars before the intense starburst during the final
merger.  However, a substantial amount of gas is left over; consequently, the
remnant is not purely stellar.  Dissipation in the gas yields a remnant with a
large, star-forming disk, owing to conservation of angular momentum.  In
Figure~\ref{fig:prj88}, we show the distribution of gas and stars in the
remnant at $t = 1.96$ Gyr. A rotationally supported gaseous disk is seen in
the merger remnant.

In Figure~\ref{fig:rem88}, we show an analysis of the stellar surface
mass density profile of the remnant, as seen when looking onto the
orbital plane.  The profile can be quite well fitted with the sum of
three physically motivated components. The first describes a central
stellar bulge formed by the starbursts, the second an exponential
stellar disk owing to ongoing star formation in the newly formed gas
disk, and the third an extended stellar spheroid originating from the
old disks destroyed during the collision. Looking at stars of
different age allows a clear identification of these components.
`Young' stars, defined here as stars forming after $T=1.2\,{\rm Gyr}$
when the merger is approximately completed, are distributed in an
exponential disk (top right panel). Note that the gas forms an
exponential disk as well, with a scale length larger by a factor 1.4,
as expected based on the Kennicutt law.  If we consider `old' stars
instead, defined here as stars forming before the first burst at
$T=0.3\,{\rm Gyr}$, we find a nearly perfect $r^{1/4}$-profile (bottom
right panel). We expect that a collisionless merger with pure stellar
disks in the colliding galaxies would exhibit a very similar final
profile. Finally, stars forming during the intense burst in the
interval $1.05\,{\rm Gyr}\le T\le 1.2\,{\rm Gyr}$ are found in a
centrally concentrated spheroid. This bulge can be fitted with a
$r^{1/4}$-profile or with an exponential, with a slight preference for
the latter.

\begin{figure}[t]
\begin{center}
\vspace*{-0.6cm}\resizebox{7.5cm}{8.0cm}{\includegraphics{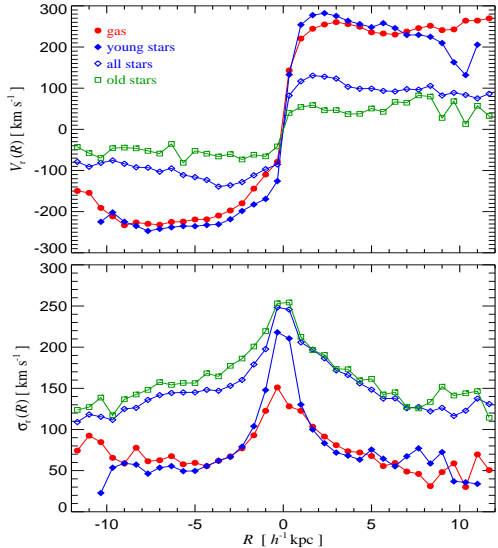}}\vspace*{-0.6cm}
\end{center}%
\caption{Kinematic profiles of the merger remnant. For a slit placed across
  the galaxy edge-on, we show the mean radial velocity (top panel), and the
  line-of-sight velocity dispersion (bottom panel). Different symbols are used
  for gas and star particles, and results for young ($T< 0.3\,{\rm Gyr}$) and
  old stars ($T>\, 1.2\,{\rm Gyr}$) are shown separately. }
\label{fig:kin88}
\end{figure}

For a decomposition of the total profile, we keep the scale lengths of
the three components identified above fixed, and only vary their
relative amplitudes. This decomposition (left panel) attributes 68\%
of the stars to the bulge component, 27\% to the stellar disk, and 5\%
to the extended spheroid.  Note that these numbers bear some
uncertainty owing to the near degeneracy of the disk and extended
spheroid profiles.  The corresponding half-mass radii are 0.3, 2.6 and
$1.7\,h^{-1}{\rm kpc}$, respectively. By coincidence, the mass of the
remaining gas is almost the same as the disk mass. Most of this
left-over gas is found in the disk, so that the stellar disk can be
expected to almost double in mass within a few Gyrs, but it will not
quite reach the mass of the bulge, unless there is cosmological infall
of fresh gas.  So the bulge, formed largely as a result of the two
bursts, is likely to remain quite prominent in this galaxy.

Further information about the structure of the remnant can be obtained from
kinematical data. In Fig.~\ref{fig:kin88}, we analyze the line-of-sight mean
velocities and velocity dispersions across a slit placed edge-on over the
remnant. Gas and young stars in the remnant are seen to be rotationally
supported, while the old stars are predominantly dispersion supported, with
small residual rotational support.  We have also directly compared the
azimuthal streaming velocities of gas and young stars with the rotation curve
measured in the plane of the disk by differentiation of the potential. These
curves agree very well, confirming the rotational support of these components.

\section{Discussion}
\label{disc}

Major mergers of disk galaxies play a prominent role in hierarchical
models of galaxy formation. They are thought to be a primary path for
the formation of large elliptical galaxies, and to give rise to
powerful starbursts and AGN accretion events. Typically,
semi-analytical models of galaxy formation make the simplifying
assumption that the gas present in a major merger is completely
consumed in a powerful burst, such that a spheroidal remnant without
a disk component is formed.

The simulation we analyzed here provides a counter-example to this
assumption, demonstrating that it cannot be correct in detail. We have
shown that gas-rich mergers can still have a significant fraction of
their gas left over after coalescence, despite the occurrence of
powerful starbursts during the merger process. As a result, the
remnant can quickly regrow a disk, such that the morphology of the
stellar remnant is never really purely spheroidal, despite being the
direct product of a major merger. This is at odds with traditional
tenets about major mergers.  The morphological evolution of galaxies
in mergers can, therefore, be more complicated than previously assumed.

While we here focused on a particular galaxy collision with a favorable
prograde orbit for disk formation, we note that our conclusions are not
restricted to this special case. However, they do depend on the modeling of
the ISM we adopted here. For simpler models of the ISM where feedback is
ignored, we are unable to simulate disk galaxies as gas-rich as the ones
considered here in a stable fashion for a sufficiently long time. Instead, the
galaxies then quickly fragment and consume most of the gas before the
collision takes place. This highlights the importance of the modeling of star
formation and feedback processes for disk stability, and for theoretical
arguments based on it \citep{Toth1992}. If disks can ``survive'' even major
mergers, they are probably less fragile than previously thought.

\acknowledgements This work was supported in part by NSF grants ACI
96-19019, AST 00-71019, AST 02-06299, and AST 03-07690, and NASA ATP
grants NAG5-12140, NAG5-13292, and NAG5-13381.  The simulations were
performed at the Center for Parallel Astrophysical Computing at
Harvard-Smithsonian Center for Astrophysics.

\bibliographystyle{apj}
\bibliography{spirals}

\end{document}